\documentclass[aps,pra,groupedaddress]{revtex4}%
\usepackage{amsmath,amsthm,amssymb}
\usepackage{graphicx}%
\usepackage{amsmath}%
\usepackage{amsfonts}%
\usepackage{amssymb}
\providecommand{\U}[1]{\protect\rule{.1in}{.1in}}
\newtheorem{criterion}[]{Criterion}
\newtheorem{prop}[]{Proposition}
\begin{document}
\title[Group structure of the IBP identities...]{Group structure of the integration-by-part identities and its application to
the reduction of multiloop integrals.}
\author{R.N. Lee}
\email{r.n.lee@inp.nsk.su}
\affiliation{The Budker Institute of Nuclear Physics and Novosibirsk State University,
Novosibirsk, Russia}

\begin{abstract}
The excessiveness of integration-by-part (IBP) identities is discussed. The
Lie-algebraic structure of the IBP identities is used to reduce the number of
the IBP equations to be considered. It is shown that Lorentz-invariance (LI)
identities do not bring any information additional to that contained in the
IBP identities, and therefore, can be discarded.

\end{abstract}
\maketitle

\section{Introduction}

Calculation of the multiloop radiative corrections in different physical
processes becomes more and more important nowadays. Mainly this is because of
the increasing precision of the modern experiments, both in high-energy
physics and in spectroscopy. The use of the IBP identities
\cite{Tkachov1981,Chetyrkin1981} is a standard approach to the effective
calculation of the loop integrals. These identities can reduce the problem of
calculation of arbitrary integral with a given topology to that of calculation
of the limited number of simpler integrals of the same topology and its subtopologies.

However, the application of the IBP identities is hampered by their infinite
number. The problem is that it is not always clear which identities should be
used to reduce a given integral. One standard approach, which has proved to be
useful, is considering the identities starting from the simplest ones and
creating a database of the rules for the reduction \cite{Laporta2000}. This
algorithm is essentially sequential, since, in order to solve the next
identity, it is necessary first to substitute all integrals which are already
in the database. Another approach to the problem is to reduce it to the
problem of \textquotedblleft division with the remainder\textquotedblright%
\ with respect to some ideal \cite{Tarasov1998a,Tarasov2004,Smirnov2006a}. The
main difficulty on this way is to derive the Gr\"{o}bner basis of the ideal
allowing for the correct determination of the simplest remainder. According to
Ref. \cite{Smirnov2007}, this procedure essentially depends on the ordering
chosen. This choice is, in some cases, not a simple problem, and the
Gr\"{o}bner basis then can be hardly found. Thus, the first approach (Laporta
approach) appears to be necessary, at least, in this situation.

One of the problems which can spoil the effectiveness of the Laporta approach
is the excessiveness of the IBP identities. When the IBP identities are
considered one-by-one, most identities do not give any additional information.
Indeed, the number of the identities grows as the volume of the region in
$\mathbb{Z}^{N}$ multiplied by the number of identities in one point, $L(L+E)$
($L$ being the number of loops, $E$, the number of external momenta), while
the number of integrals involved grows only as the volume of the region in
$\mathbb{Z}^{N}$. Thus, in the asymptotics, only one identity out of $L(L+E)$
give new information. Other identities in the Laporta approach are checked to
reduce to $0=0$. Unfortunately, this check can be a very time consuming
calculation. The determination of the minimal set of the IBP identities is
also important for the analytical solution of identities (i.e., the derivation
of the reduction rules) as it allows one to consider a smaller set of the identities.

Though the algebraic manipulation with the IBP identities has been used for a
long time, the observation that IBP identities form a closed Lie algebra has
not been exploited so far. The main purpose of this paper is to demonstrate
how the Lie-algebraic properties of the IBP equations can be used for both
{}``division with the remainder'' algorithms and Laporta-like algorithms.

\section{General discussion}

Assume that we are interested in the calculation of the $L$-loop integral
depending on the $E$ external momenta $p_{1},\ldots,p_{E}$. There are $N$
scalar products depending on the loop momenta $l_{i}$:%

\begin{align}
s_{ik}  &  =l_{i}\cdot q_{k}\,,\ 1\leqslant i\leqslant L,k\leqslant
L+E,\nonumber\\
N  &  =L(L+1)/2+LE
\end{align}
where $q_{1,\ldots,L}=l_{1,\ldots,L}$, $q_{L+1,\ldots,L+E}=p_{1,\ldots,E}$.

The loop integral has the form%
\begin{equation}
J\left(  \boldsymbol{n}\right)  =J(n_{1},n_{2},\ldots,n_{N})=\int
d^{\mathcal{D}}l_{1}\ldots d^{\mathcal{D}}l_{L}j(\boldsymbol{n})=\int
\frac{d^{\mathcal{D}}l_{1}\ldots d^{\mathcal{D}}l_{L}}{D_{1}^{n_{1}}%
D_{2}^{n_{2}}\ldots D_{N}^{n_{N}}} \label{eq:J}%
\end{equation}
where the scalar functions $D_{\alpha}$ are linear polynomials with respect to
$s_{ij}$. The functions $D_{\alpha}$ are assumed to be linearly independent
and to form a complete basis in the sense that any non-zero linear combination
of them depends on the loop momenta, and any $s_{ik}$ can be expressed in
terms of $D_{\alpha}$. Thus, each integral is associated with a point in
$\mathbb{Z}^{N}$. Some of the functions $D_{\alpha}$ correspond to the
denominators of the propagators, the other correspond to the irreducible
numerators. E.g., the $K$-legged $L$-loop diagram corresponds to $E=K-1$ and
the maximal number of denominators is $M=E+3L-2$, so that the rest
$N-M=(L-1)(L+2E-4)/2$ functions correspond to irreducible numerators. For
vacuum diagrams, $M=3(L-1)$, and $N-M=(L-2)(L-3)/2$.

The IBP identities are based on the fact that, in the dimensional
regularization, the integral of the total derivative is zero. They are derived
from the identity%
\begin{equation}
0=\int d^{\mathcal{D}}l_{1}\ldots d^{\mathcal{D}}l_{L}O_{ik}j(\boldsymbol{n}%
)=\int d^{\mathcal{D}}l_{1}\ldots d^{\mathcal{D}}l_{L}\frac{\partial}{\partial
l_{i}}\cdot q_{k}j(\boldsymbol{n})\,.
\end{equation}

Performing the differentiation in the right-hand side and expressing the
scalar products via $D_{\alpha}$, we obtain the recurrence relation for the
function $J$.

There is also another class of identities, called Lorentz-invariance (LI)
identities due to the fact that the integral (\ref{eq:J}) is Lorentz scalar
\cite{Gehrmann2000}. They have the form%
\begin{equation}
p_{i}^{\mu}p_{j}^{\nu}\left(  \sum_{k}p_{k[\nu}\frac{\partial}{\partial
p_{k}^{\mu]}}\right)  J(n_{1},n_{2},\ldots,n_{N})=0 \label{eq:LI}%
\end{equation}

The differential operator in braces is nothing but the generator of the
Lorentz transformation in the linear space of scalar functions depending on
$p_{k}$. If we explicitly act by the differential operator on the integrand,
we obtain LI identity. Though these identities can be convenient in some
cases, they can be easily represented as some linear combination of the IBP
identities (see Appendix) and will not be considered in the following.

For the reduction procedure to work, it is necessary to define some suitable
ordering of the integrals, i.e., the ordering in $\mathbb{Z}^{N}$. First, one
introduces the notion of sectors in $\mathbb{Z}^{N}$. The $(\theta_{1}%
,\ldots,\theta_{N})$ sector, where $\theta_{i}=0,1$, is a set of all points
$\left(  n_{1},\ldots,n_{N}\right)  $ in $\mathbb{Z}^{N}$ whose coordinates
obey the condition
\begin{align}
\mathrm{sign}\left(  n_{\alpha}-1/2\right)   &  = 2\theta_{\alpha}-1.
\end{align}

In particular, the point $(\theta_{1},\ldots,\theta_{N})$ belongs to the
$(\theta_{1},\ldots,\theta_{N})$ sector, and can be referred to as the
\emph{corner point of the sector}. Owing to this definition, the integrals of
the same sector have the same number of denominators. It is natural to
consider the integrals with less denominators to be simpler. When the number
of denominators coincides, we will consider the integrals with smaller total
power of the numerators and denominators to be simpler. Then goes the number
of the numerators and the last is the lexicographical ordering. Thus, two
points $\boldsymbol{n}=\left(  n_{1},n_{2},\ldots\right)  $ and
$\boldsymbol{n}^{\prime}=\left(  n_{1}^{\prime},n_{2}^{\prime},\ldots\right)
$, are said to be ordered as $\boldsymbol{n}\prec\boldsymbol{n}^{\prime}$ iff
there exists $i_{0},-2<i_{0}\leqslant N$, such that $n_{i_{0}}<n_{i_{0}%
}^{\prime}$and for any $i,-2\leqslant i<i_{0}$ holds $n_{i}=n_{i}^{\prime}$.
Here $n_{-2},n_{-1}$, and $n_{0}\ $are determined as%

\begin{equation}
n_{-2}=\sum_{\alpha=1}^{N}\Theta\left(  n_{\alpha}-1/2\right)  =\sum
_{\alpha=1}^{N}\theta_{\alpha},\quad n_{-1}=\sum_{\alpha=1}^{N}|n_{\alpha
}|,\quad n_{0}=\sum_{\alpha=1}^{N}\Theta\left(  -n_{\alpha}+1/2\right)  .
\end{equation}
The integral $J\left(  \boldsymbol{n}\right)  $ is considered to be simpler
than $J\left(  \boldsymbol{n}^{\prime}\right)  $ if $\boldsymbol{n}%
\prec\boldsymbol{n}^{\prime}$. According to this ordering, the integral
$J(\theta_{1},\ldots,\theta_{N})$ is the simplest integral of $(\theta
_{1},\ldots,\theta_{N})$ sector.

\section{Operator representation}

Let us introduce, similar to Ref. \cite{Smirnov2006a}, the operators
$A_{\alpha}$ and $B_{\alpha}$ acting on functions in $\mathbb{Z}^{N}$ as
follows
\begin{align}
\left(  A_{\alpha}f\right)  \left(  n_{1},\ldots,n_{N}\right)   &  =n_{\alpha
}\,f\left(  n_{1},\ldots,n_{\alpha}+1,\ldots,n_{N}\right)  ,\nonumber\\
\left(  B_{\alpha}f\right)  \left(  n_{1},\ldots,n_{N}\right)   &  =f\left(
n_{1},\ldots,n_{\alpha}-1,\ldots,n_{N}\right)  .
\end{align}
Note that these operators act on function, but not on its arguments, and
should not be confused with the conventional $\mathbf{n}^{\pm}$ index shifting
operators. Using these operators, we can express the IBP identities as
constraints on the function $J$ having the form%
\begin{align*}
-PJ  &  =0,\\
P  &  =a^{\alpha\beta}A_{\alpha}B_{\beta}+b^{\alpha}A_{\alpha}+c,
\end{align*}
where $a^{\alpha\beta},\,b^{\alpha},\,c$ are some coefficients. We will denote
the operator, corresponding to the $O_{ik}$ as $P_{ik}$:
\begin{equation}
-\left(  P_{ik}J\right)  \left(  \boldsymbol{n}\right)  =\int d^{\mathcal{D}%
}l_{1}\ldots d^{\mathcal{D}}l_{L}O_{ik}j\left(  \boldsymbol{n}\right)  .
\end{equation}
Note that the operators $A_{\alpha},B_{\alpha}$ form Weyl algebra,%
\begin{equation}
\lbrack A_{\alpha},B_{\beta}]=\delta_{\alpha\beta}.
\end{equation}

Let $\mathcal{L}$ be the left ideal generated by operators $P_{ik}$, i.e. a
set, consisting of all operators, which can be represented as%
\begin{equation}
\sum_{i,k}C_{ik}P_{ik},
\end{equation}
where $C_{ik}$ are some polynomials of $A_{1},\ldots A_{N},B_{1},\ldots,B_{N}%
$. This ideal has a simple meaning: for any $L\in\mathcal{L}$ the relation%
\begin{equation}
\left(  LJ\right)  \left(  n_{1},\ldots,n_{N}\right)  =0
\end{equation}
is a linear combination of some IBP identities. In fact, any linear
combination of the IBP identities can be represented in a more specific form%
\begin{equation}
\left(  LJ\right)  \left(  1,\ldots,1\right)  =0,
\end{equation}
since shifting of the indices can be done by acting from the left with some
powers of $A_{\alpha}$ or $B_{\alpha}$. At first glance, the problem of
reduction is equivalent to that of division with the remainder by the ideal
$\mathcal{L}$, which is effectively solved by the construction of the
Gr\"{o}bner basis. However, there is an additional obstacle. Note that for any
function $f$ of $N$ integer variables the following relation holds%
\begin{equation}
\left(  B_{\alpha}A_{\alpha}f\right)  \left(  1,\ldots,1\right)  =0\,.\text{
(no summation)} \label{eq:Right}%
\end{equation}
Indeed,
\begin{equation}
\left(  B_{\alpha}A_{\alpha}f\right)  \left(  1,\ldots,1\right)  =\left(
A_{\alpha}f\right)  \left(  1,\ldots,\overset{\alpha}{0},\ldots,1\right)
=0\,\times f\left(  1,\ldots,\overset{\alpha}{1},\ldots,1\right)  =0,
\end{equation}
where the overscript $\alpha$ denotes the position of the index. Let
$\mathcal{R}$ be the right ideal generated by the elements $\left(  B_{1}%
A_{1}\right)  ,\ldots,\left(  B_{N}A_{N}\right)  $. By definition, it consists
of all operators of the form%
\begin{equation}
R=\sum_{\alpha}B_{\alpha}A_{\alpha}C_{\alpha},
\end{equation}
where $C_{\alpha}$ are some polynomials of $A_{1},\ldots A_{N},B_{1}%
,\ldots,B_{N}$. It follows from Eq. (\ref{eq:Right}) that%
\begin{equation}
\left(  Rf\right)  \left(  1,\ldots,1\right)  =0.\,
\end{equation}
Thus, for the reduction procedure to work, we have to have an algorithm of
division with the remainder by the direct sum of the left ideal $\mathcal{L}$
and the right ideal $\mathcal{R}$. That means that we have to invent the
algorithm allowing the decomposition%
\begin{equation}
p=L+R+r,
\end{equation}
where $L\in\mathcal{L},\,R\in\mathcal{R}$, and $r$ is the simplest possible
with respect to the ordering chosen. Even though the problem is clearly
formulated, such algorithm appears to be unknown so far.

\section{Lie-algebraic structure of the IBP identities}

The operators
\begin{equation}
O_{ik}=\frac{\partial}{\partial l_{i}}\cdot q_{k}%
\end{equation}
form a closed algebra with the commutation relations%
\begin{equation}
\left[  O_{ik},O_{jl}\right]  =\delta_{il}O_{jk}-\delta_{jk}O_{il}.
\label{eq:CommutationRelations}%
\end{equation}
We can easily check that the operators $P_{ik}$ obey the same commutation
relations as $O_{ik}$. This algebra is nothing but the algebra of the group of
linear changes of variables%
\begin{equation}
l_{i}\rightarrow M_{ik}q_{k}. \label{eq:LinearTransformation}%
\end{equation}
The operator $O_{ik}$ corresponds to the infinitesimal transformation
$l_{i}\rightarrow l_{i}^{\prime}=l_{i}+\epsilon q_{k}$ in the sense that
\begin{equation}
f(s_{lm}^{\prime})d^{\mathcal{D}}l_{1}^{\prime}\ldots d^{\mathcal{D}}%
l_{L}^{\prime}=\left\{  f(s_{lm})+\epsilon\left[  \frac{\partial}{\partial
l_{i}}\cdot q_{k}f(s_{lm})\right]  \right\}  d^{\mathcal{D}}l_{1}\ldots
d^{\mathcal{D}}l_{L}+O(\epsilon^{2}).
\end{equation}
Note that the so-called symmetry relations are also the consequences of the
invariance of the integrand under the action of some elements of this group.

The scaleless integral can be defined as the one which gains additional
non-unity factor under some transformation (\ref{eq:LinearTransformation}).
This definition corresponds to the conventional notion of scaleless integrals.
E.g., owing to this definition, the integral
\begin{equation}
I=\int\frac{d^{\mathcal{D}}l_{1}\,d^{\mathcal{D}}l_{2}}{\left(  l_{1}%
-p\right)  ^{2}l_{2}^{2}\left(  l_{1}-p-l_{2}\right)  ^{2}}%
\end{equation}
is scalelless as, under the transformation%
\begin{align}
l_{1}  &  \rightarrow\alpha l_{1}+\left(  1-\alpha\right)  p,\nonumber\\
l_{2}  &  \rightarrow\alpha l_{2}\,,
\end{align}
it transforms as%
\begin{equation}
I\rightarrow\alpha^{2\mathcal{D}-6}\,I.
\end{equation}
In dimensional regularization, the scaleless integrals are zero, as well as
the integrals which differ from the scaleless ones by additional polynomial
factor in the numerator. Thus, once the integral in the corner point of the
sector is scaleless, the whole sector is zero. Let us prove a simple criterion
of zero sectors.

\begin{criterion}
If the solution of all IBP relations in the corner point of the sector
$(\theta_{1},\ldots,\theta_{N})$ results in the identity
\begin{equation}
J(\theta_{1},\ldots,\theta_{N})=0, \label{eq:j1}%
\end{equation}

then this sector is zero, i.e., all integrals of this sector are zero.
\end{criterion}

Indeed, by the condition, $j(\theta_{1},\ldots,\theta_{N})$ can be represented
as the action of some linear combination of $O_{ik}$ on $j(\theta_{1}%
,\ldots,\theta_{N})$. Since these operators are generators of the
transformation (\ref{eq:LinearTransformation}), we can conclude that
$J(\theta_{1},\ldots,\theta_{N})$ is scaleless and thus, the whole sector
$(\theta_{1},\ldots,\theta_{N})$ is zero. This criterion gives a simple and
convenient way to determine zero sectors.

\section{Excessiveness of the IBP identities set.}

In this Section we describe some consequences of the algebraic structure of
the IBP identities.

\begin{prop}
Let $L\geqslant2$. Then of all $L(L+E)$ IBP identities we can consider only
identities, generated by the operators:%
\begin{align}
\frac{\partial}{\partial l_{i}}\cdot l_{i+1},  &  \quad i=1,\ldots,L,\quad
l_{L+1}\equiv l_{1}\nonumber\\
\frac{\partial}{\partial l_{1}}\cdot p_{j},  &  \quad j=1,\ldots,E\nonumber\\
\sum_{i=1}^{L}  &  \frac{\partial}{\partial l_{i}}\cdot l_{i}
\label{eq:reducedset}%
\end{align}

\end{prop}

Indeed, this set of operators form the multiplicative basis of the Lie-algebra
(\ref{eq:CommutationRelations}), i.e., the rest of the operators can be
obtained from the commutators of the chosen ones. The total number of the
operators in the set (\ref{eq:reducedset}) is $L+E+1$, which is smaller than
the original $L(L+E)$ for $L\geqslant2$. This simple fact can be used for the
construction of the reduction rules and also for the Laporta algorithm.
Nevertheless, such system of the IBP identities is still overdetermined. In
the asymptotics only $1/(L+E+1)$ part of the identities gives new information.

Now we prove a more refined criterion for the identities which can be thrown
away without loss of information. Let $S=\{P_{1},P_{2},\ldots,P_{K}\}$ be some
set of the IBP operators, and $P$ be some IBP operator with the following
property: its commutator with any $P_{k}\in S$ is a linear combination of
$P_{i}\in S$. Then we have the following

\begin{criterion}
\label{crit:2}If for some point $\boldsymbol{n}\in\mathbb{Z}^{N}$the integral
$J(\boldsymbol{n})$ can be expressed via simpler integrals with the help of
the identities obtained from the operators in $S$, then the identity
$(PJ)(\boldsymbol{n})=0$ can be represented as a linear combination of the
identities obtained from the operators in $S$ and the identities of the form
$(PJ)(\boldsymbol{n}^{\prime})=0$ with $\boldsymbol{n}^{\prime}\prec
\boldsymbol{n}$.
\end{criterion}

\begin{proof}
To prove it, we note that, since the integral $J(\boldsymbol{n})$ can be expressed via simpler integrals with the help of the identities
obtained from the operators in $S$, there exist such polynomials $C_{l}$ that for any function $f$
\begin{equation}
\sum_{l}(C_{l}P_{l}f)(\boldsymbol{1})=f(\boldsymbol{n})+o\left(\boldsymbol{n}\right),\end{equation}
$o\left(\boldsymbol{n}\right)$ denotes here the linear combination of $f(\boldsymbol{n}^{\prime})$ with
$\boldsymbol{n}^{\prime}\prec\boldsymbol{n}$. Now we substitute $f=PJ$:
\begin{equation}
\sum_{l}(C_{l}P_{l}PJ)(\boldsymbol{1})=PJ(\boldsymbol{n})+Po\left(\boldsymbol{n}\right),\end{equation}
and use the commutation relation\begin{equation} P_{l}P=PP_{l}+\sum_{m}c_{lm}P_{m}\end{equation} where $c_{lm}$ are some constants.
We obtain\begin{equation}
PJ(\boldsymbol{n})=Po\left(\boldsymbol{n}\right)+\sum_{l}(C_{l}PP_{l}J)(\boldsymbol{1})+\sum_{lm}(C_{l}c_{lm}P_{m}J)(\boldsymbol{1})\end{equation}
Here $Po\left(\boldsymbol{n}\right)$ denotes the linear combination
of $PJ(\boldsymbol{n}^{\prime})$ with $\boldsymbol{n}^{\prime}\prec\boldsymbol{n}$, and the last two terms is a linear combination of of the
identities obtained from the operators in $S$.
\end{proof}
Let us consider the sequence of the operators
\begin{equation}
\left\{  \mathcal{P}_{1},\ldots\mathcal{P}_{N}\right\}  =\left\{
P_{1,L+E},\ldots,P_{1,1},P_{2,L+E},\ldots,P_{2,2},\ldots,P_{L,L+E}%
,\ldots,P_{L,L}\right\}  \label{eq:seq}%
\end{equation}
Note that the number of the operators in this sequence equals to
$N=L(L+1)/2+LE$, the total number of the denominators and numerators in basis.
This sequence has the following property: for any $\mathcal{P}_{i}$ the
criterion \ref{crit:3} applies with $S=\left\{  \mathcal{P}_{1},\ldots
\mathcal{P}_{i-1}\right\}  $. Suppose we have solved the identities, generated
by operators $\left\{  \mathcal{P}_{1},\ldots\mathcal{P}_{i-1}\right\}  $,
then the identities, generated by $\mathcal{P}_{i}$ should be solved only in
points for which there are no reduction rules yet. On each step of this
procedure the {}``dimension'' of the set of such points is decreased by one,
thus, when we have considered all operators in the sequence, we have only
finite number of the integrals, which are not yet reduced. The rest $L(L-1)/2$
IBP identities can be used for the reduction of integrals in these points.
Note, that the mean number of the identities, considered in each point is $1$,
as it should be. The choice of the sequence (\ref{eq:seq}) is, of course, not unique.

Now we derive the criterion which can be used for the algorithm combining the
{}``division with the remainder{}`` and Laporta method.

\begin{criterion}
\label{crit:3}Let in some sector the identities generated by some operator
$P=\sum c_{ik}P_{ik}$ have the form%
\begin{equation}
(PJ)(\boldsymbol{n})=J(\tilde{\boldsymbol{n}})+o\left(  \tilde{\boldsymbol{n}%
}\right)  \label{eq:identityform}%
\end{equation}
and for any $\boldsymbol{n}$ in the sector holds%
\begin{equation}
\boldsymbol{n}\prec\tilde{\boldsymbol{n}}.
\end{equation}

Then the identity generated by another operator $P^{\prime}=\sum
c_{ik}^{\prime}P_{ik}\neq P$ in point $\tilde{\boldsymbol{n}}$, corresponding
to the integral expressed by $P$ , can be represented as a linear combination
of the identities of the form $PJ$ and the identities in simpler points.
\end{criterion}

\begin{proof}
To prove it, let us consider the identity\begin{equation}
\left(PP^{\prime}J\right)\left(\boldsymbol{n}\right)=\left(P^{\prime}J\right)\left(\tilde{\boldsymbol{n}}\right)+\left(P^{\prime}o\right)\left(\tilde{\boldsymbol{n}}\right)=P^{\prime}PJ\left(\boldsymbol{n}\right)+\left(\left[P,P^{\prime}\right]J\right)\left(\boldsymbol{n}\right).\end{equation}
Thus \begin{equation}
\left(P^{\prime}J\right)\left(\tilde{\boldsymbol{n}}\right)=\left(P^{\prime}PJ\right)\left(\boldsymbol{n}\right)-\left(P^{\prime}o\right)\left(\tilde{\boldsymbol{n}}\right)+\left(\left[P,P^{\prime}\right]J\right)\left(\boldsymbol{n}\right).\end{equation}
The first term in the left-hand side is some linear combination of the IBP identities generated by $P$, and the last two terms contain only
identities in the points $\boldsymbol{n}^{\prime}\prec\tilde{\boldsymbol{n}}$. Thus, the identity $P^{\prime}J\left(\tilde{\boldsymbol{n}}\right)$
is dependent on the identities of the form $PJ$ and on the identities in simpler points.
\end{proof}

The basic idea of application of this criterion is the following. Consider the
IBP identities in general point of the sector. We can either use the set of
identities, generated by $P_{ik}$ or by other $L(L+E)$ independent linear
combinations of $P_{ik}$. In fact, it is natural to pass to such linear
combinations, in which all most complex integrals are different ("solve" the
identities in general point). Among these identities, select one, having the
form (\ref{eq:identityform}) with $\boldsymbol{n}\prec\tilde{\boldsymbol{n}}$
and $\tilde{\boldsymbol{n}}$ corresponding to the integral not yet expressed.
Determine the points of the sector corresponding to the integrals for which
this identity does not work. Consider the other identities only in these
points. Repeate the same steps, starting from the selection of the identity.
After some iterations, we might be unable to find the identity matching the
conditions. At this stage, we have to solve the rest of the identities only in
the points, corresponding to the integrals, which are not yet expressed. E.g.,
we can use the Laporta method. The advantage of this combined approach is that
the {}\textquotedblleft dimension\textquotedblright\ of the set of points in
which we use the Laporta method is usually essentially less than $N$.

\section{Conclusion}

In the present paper, we have considered the dependencies between the IBP
identities. Account of these dependencies dramatically decreases the number of
the identities to be considered. They come from the fact that the
corresponding operators $P_{ik}$ form a closed Lie algebra of the group of
linear change of variables. Using this interpretation of the IBP identities,
we have proved the simple criterion of the zero sectors. The two criteria of
the excessiveness of an identity were proven. Probably, using the Criterion
\ref{crit:2}, it is possible to prove the finiteness of the number of master
integrals in general case. Indeed, selecting the sequence of the operators as
it was described, we decrease on each step the ``dimension'' of the set of the
unexpressed integrals by one. Since the number of the operators in this
sequence equals $N$, we are left with the set of ``dimension'' zero, i.e.,
consisting of finite number of points. The Criterion \ref{crit:3} can be used
for the combined algorithms in which the Laporta reduction is performed on
some subset of the points of the given sector with the dimension less than
$N$. It was also shown that the Lorentz-invariance identities can be
completely discarded.

This work was supported by RFBR Grant No. 07-02-00953. I also thank for warm
hospitality the Max-Planck Institute for Quantum Optics, Garching, where a
part of this work was done.

\appendix

\section{Expressing LI identities via IBP identities.}

In this Appendix we will show that the LI identities always can be represented
as a linear combination of the IBP identities.

Let us note that the integrand $j$ in Eq. (\ref{eq:J}) is a scalar function of
$q_{i}$. Thus, the operator%
\begin{equation}
\sum_{k=1}^{L+E}q_{k[\nu}\frac{\partial}{\partial q_{k}^{\mu]}}=\sum_{k=1}%
^{L}l_{k[\nu}\frac{\partial}{\partial l_{k}^{\mu]}}+\sum_{k=1}^{E}p_{k[\nu
}\frac{\partial}{\partial p_{k}^{\mu]}},
\end{equation}

when acting on the integrand $j$, annihilates it identically. Indeed, this
operator is nothing, but the generator of Lorentz transformations in the
linear space of functions of $q_{i}$. Thus, the operator in Eq. (\ref{eq:LI}),
when acting on the integrand, can be represented as follows:%
\begin{align*}
p_{i}^{\mu}p_{j}^{\nu}\sum_{k=1}^{E}p_{k[\mu}\frac{\partial}{\partial
p_{k}^{\nu]}}  &  j= p_{i}^{\mu}p_{j}^{\nu}\sum_{k=1}^{L}l_{k[\nu}%
\frac{\partial}{\partial l_{k}^{\mu]}}j-p_{i}^{\mu}p_{j}^{\nu}\sum_{k=1}%
^{L+E}q_{k[\nu}\frac{\partial}{\partial q_{k}^{\mu]}}j\\
&  = p_{i}^{\mu}p_{j}^{\nu}\sum_{k=1}^{L}l_{k[\nu}\frac{\partial}{\partial
l_{k}^{\mu]}}j\\
&  = \sum_{k=1}^{L}\left[  \left(  p_{i}\cdot l_{k}\right)  p_{j}\cdot
\frac{\partial}{\partial l_{k}}-\left(  p_{j}\cdot l_{k}\right)  p_{i}%
\cdot\frac{\partial}{\partial l_{k}}\right]  j\\
&  = \sum_{k=1}^{L}\left[  \frac{\partial}{\partial l_{k}}\cdot p_{j}\left(
p_{i}\cdot l_{k}\right)  -\frac{\partial}{\partial l_{k}}\cdot p_{i}\left(
p_{j}\cdot l_{k}\right)  \right]  j
\end{align*}

Taking into account that the scalar products $\left(  p_{i,j}\cdot
l_{k}\right)  $ in the last line can be expressed via $D_{\alpha}$, we
conclude, that any LI identity can be represented as a linear combination of
the IBP identities.

%\bibliographystyle{plain}
%\bibliography{refs}

\end{document}